\documentclass[aps,prd,reprint,superscriptaddress]{revtex4-1}

\usepackage{amsmath} 
\usepackage{multirow}
\usepackage{graphicx}
\usepackage{epstopdf}
\usepackage{xspace} 

\usepackage{scalerel}
\newlength{\heightm}
\AtBeginDocument{\settoheight{\heightm}{$m$}}
\newcommand{\dmsqpbar}{\Delta\llap{\raisebox{\heightm+0.5pt}{\scaleobj{0.3}{\hstretch{1.3}(}}}  \overline{m}
\rlap{\raisebox{\heightm+0.5pt}{\scaleobj{0.3}{\hstretch{1.3})}}} \vphantom{m}^{\hphantom{\scaleobj{0.4}{)}}2}}
\newcommand{\dmsqbar}{\Delta\overline{m}^2} 
\newcommand{\dmsq}{\Delta m^2} 

\newlength{\heighttheta}
\AtBeginDocument{\settoheight{\heighttheta}{$\theta$}}
\newcommand{\thetapbar}{\llap{\raisebox{\heighttheta+0.5pt}{\scaleobj{0.3}{\hstretch{1.3}(}}}  \overline{\theta}\rlap{\raisebox{\heighttheta+0.5pt}{\scaleobj{0.3}{\hstretch{1.3})}}}}
\newcommand{\thetabar}{\overline{\theta}}

\newlength{\heightnu}
\AtBeginDocument{\settoheight{\heightnu}{$\nu$}}

\begin{document}

\title{Updated T2K measurements of muon neutrino and antineutrino disappearance using $1.5 \times 10^{21}$ protons on target}



\newcommand{\INSTEE}{\affiliation{University of Bern, Albert Einstein Center for Fundamental Physics, Laboratory for High Energy Physics (LHEP), Bern, Switzerland}}
\newcommand{\INSTFE}{\affiliation{Boston University, Department of Physics, Boston, Massachusetts, U.S.A.}}
\newcommand{\INSTD}{\affiliation{University of British Columbia, Department of Physics and Astronomy, Vancouver, British Columbia, Canada}}
\newcommand{\INSTGA}{\affiliation{University of California, Irvine, Department of Physics and Astronomy, Irvine, California, U.S.A.}}
\newcommand{\INSTI}{\affiliation{IRFU, CEA Saclay, Gif-sur-Yvette, France}}
\newcommand{\INSTGB}{\affiliation{University of Colorado at Boulder, Department of Physics, Boulder, Colorado, U.S.A.}}
\newcommand{\INSTFG}{\affiliation{Colorado State University, Department of Physics, Fort Collins, Colorado, U.S.A.}}
\newcommand{\INSTFH}{\affiliation{Duke University, Department of Physics, Durham, North Carolina, U.S.A.}}
\newcommand{\INSTBA}{\affiliation{Ecole Polytechnique, IN2P3-CNRS, Laboratoire Leprince-Ringuet, Palaiseau, France }}
\newcommand{\INSTEF}{\affiliation{ETH Zurich, Institute for Particle Physics, Zurich, Switzerland}}
\newcommand{\INSTEG}{\affiliation{University of Geneva, Section de Physique, DPNC, Geneva, Switzerland}}
\newcommand{\INSTDG}{\affiliation{H. Niewodniczanski Institute of Nuclear Physics PAN, Cracow, Poland}}
\newcommand{\INSTCB}{\affiliation{High Energy Accelerator Research Organization (KEK), Tsukuba, Ibaraki, Japan}}
\newcommand{\INSTED}{\affiliation{Institut de Fisica d'Altes Energies (IFAE), The Barcelona Institute of Science and Technology, Campus UAB, Bellaterra (Barcelona) Spain}}
\newcommand{\INSTEC}{\affiliation{IFIC (CSIC \& University of Valencia), Valencia, Spain}}
\newcommand{\INSTEI}{\affiliation{Imperial College London, Department of Physics, London, United Kingdom}}
\newcommand{\INSTGF}{\affiliation{INFN Sezione di Bari and Universit\`a e Politecnico di Bari, Dipartimento Interuniversitario di Fisica, Bari, Italy}}
\newcommand{\INSTBE}{\affiliation{INFN Sezione di Napoli and Universit\`a di Napoli, Dipartimento di Fisica, Napoli, Italy}}
\newcommand{\INSTBF}{\affiliation{INFN Sezione di Padova and Universit\`a di Padova, Dipartimento di Fisica, Padova, Italy}}
\newcommand{\INSTBD}{\affiliation{INFN Sezione di Roma and Universit\`a di Roma ``La Sapienza'', Roma, Italy}}
\newcommand{\INSTEB}{\affiliation{Institute for Nuclear Research of the Russian Academy of Sciences, Moscow, Russia}}
\newcommand{\INSTHA}{\affiliation{Kavli Institute for the Physics and Mathematics of the Universe (WPI), The University of Tokyo Institutes for Advanced Study, University of Tokyo, Kashiwa, Chiba, Japan}}
\newcommand{\INSTCC}{\affiliation{Kobe University, Kobe, Japan}}
\newcommand{\INSTCD}{\affiliation{Kyoto University, Department of Physics, Kyoto, Japan}}
\newcommand{\INSTEJ}{\affiliation{Lancaster University, Physics Department, Lancaster, United Kingdom}}
\newcommand{\INSTFC}{\affiliation{University of Liverpool, Department of Physics, Liverpool, United Kingdom}}
\newcommand{\INSTFI}{\affiliation{Louisiana State University, Department of Physics and Astronomy, Baton Rouge, Louisiana, U.S.A.}}
\newcommand{\INSTJ}{\affiliation{Universit\'e de Lyon, Universit\'e Claude Bernard Lyon 1, IPN Lyon (IN2P3), Villeurbanne, France}}
\newcommand{\INSTHB}{\affiliation{Michigan State University, Department of Physics and Astronomy,  East Lansing, Michigan, U.S.A.}}
\newcommand{\INSTCE}{\affiliation{Miyagi University of Education, Department of Physics, Sendai, Japan}}
\newcommand{\INSTDF}{\affiliation{National Centre for Nuclear Research, Warsaw, Poland}}
\newcommand{\INSTFJ}{\affiliation{State University of New York at Stony Brook, Department of Physics and Astronomy, Stony Brook, New York, U.S.A.}}
\newcommand{\INSTGJ}{\affiliation{Okayama University, Department of Physics, Okayama, Japan}}
\newcommand{\INSTCF}{\affiliation{Osaka City University, Department of Physics, Osaka, Japan}}
\newcommand{\INSTGG}{\affiliation{Oxford University, Department of Physics, Oxford, United Kingdom}}
\newcommand{\INSTBB}{\affiliation{UPMC, Universit\'e Paris Diderot, CNRS/IN2P3, Laboratoire de Physique Nucl\'eaire et de Hautes Energies (LPNHE), Paris, France}}
\newcommand{\INSTGC}{\affiliation{University of Pittsburgh, Department of Physics and Astronomy, Pittsburgh, Pennsylvania, U.S.A.}}
\newcommand{\INSTFA}{\affiliation{Queen Mary University of London, School of Physics and Astronomy, London, United Kingdom}}
\newcommand{\INSTE}{\affiliation{University of Regina, Department of Physics, Regina, Saskatchewan, Canada}}
\newcommand{\INSTGD}{\affiliation{University of Rochester, Department of Physics and Astronomy, Rochester, New York, U.S.A.}}
\newcommand{\INSTHC}{\affiliation{Royal Holloway University of London, Department of Physics, Egham, Surrey, United Kingdom}}
\newcommand{\INSTBC}{\affiliation{RWTH Aachen University, III. Physikalisches Institut, Aachen, Germany}}
\newcommand{\INSTFB}{\affiliation{University of Sheffield, Department of Physics and Astronomy, Sheffield, United Kingdom}}
\newcommand{\INSTDI}{\affiliation{University of Silesia, Institute of Physics, Katowice, Poland}}
\newcommand{\INSTEH}{\affiliation{STFC, Rutherford Appleton Laboratory, Harwell Oxford,  and  Daresbury Laboratory, Warrington, United Kingdom}}
\newcommand{\INSTCH}{\affiliation{University of Tokyo, Department of Physics, Tokyo, Japan}}
\newcommand{\INSTBJ}{\affiliation{University of Tokyo, Institute for Cosmic Ray Research, Kamioka Observatory, Kamioka, Japan}}
\newcommand{\INSTCG}{\affiliation{University of Tokyo, Institute for Cosmic Ray Research, Research Center for Cosmic Neutrinos, Kashiwa, Japan}}
\newcommand{\INSTGI}{\affiliation{Tokyo Metropolitan University, Department of Physics, Tokyo, Japan}}
\newcommand{\INSTF}{\affiliation{University of Toronto, Department of Physics, Toronto, Ontario, Canada}}
\newcommand{\INSTB}{\affiliation{TRIUMF, Vancouver, British Columbia, Canada}}
\newcommand{\INSTG}{\affiliation{University of Victoria, Department of Physics and Astronomy, Victoria, British Columbia, Canada}}
\newcommand{\INSTDJ}{\affiliation{University of Warsaw, Faculty of Physics, Warsaw, Poland}}
\newcommand{\INSTDH}{\affiliation{Warsaw University of Technology, Institute of Radioelectronics, Warsaw, Poland}}
\newcommand{\INSTFD}{\affiliation{University of Warwick, Department of Physics, Coventry, United Kingdom}}
\newcommand{\INSTGE}{\affiliation{University of Washington, Department of Physics, Seattle, Washington, U.S.A.}}
\newcommand{\INSTGH}{\affiliation{University of Winnipeg, Department of Physics, Winnipeg, Manitoba, Canada}}
\newcommand{\INSTEA}{\affiliation{Wroclaw University, Faculty of Physics and Astronomy, Wroclaw, Poland}}
\newcommand{\INSTHE}{\affiliation{Yokohama National University, Faculty of Engineering, Yokohama, Japan}}
\newcommand{\INSTH}{\affiliation{York University, Department of Physics and Astronomy, Toronto, Ontario, Canada}}

\INSTEE
\INSTFE
\INSTD
\INSTGA
\INSTI
\INSTGB
\INSTFG
\INSTFH
\INSTBA
\INSTEF
\INSTEG
\INSTDG
\INSTCB
\INSTED
\INSTEC
\INSTEI
\INSTGF
\INSTBE
\INSTBF
\INSTBD
\INSTEB
\INSTHA
\INSTCC
\INSTCD
\INSTEJ
\INSTFC
\INSTFI
\INSTJ
\INSTHB
\INSTCE
\INSTDF
\INSTFJ
\INSTGJ
\INSTCF
\INSTGG
\INSTBB
\INSTGC
\INSTFA
\INSTE
\INSTGD
\INSTHC
\INSTBC
\INSTFB
\INSTDI
\INSTEH
\INSTCH
\INSTBJ
\INSTCG
\INSTGI
\INSTF
\INSTB
\INSTG
\INSTDJ
\INSTDH
\INSTFD
\INSTGE
\INSTGH
\INSTEA
\INSTHE
\INSTH

\author{K.\,Abe}\INSTBJ
\author{J.\,Amey}\INSTEI
\author{C.\,Andreopoulos}\INSTEH\INSTFC
\author{M.\,Antonova}\INSTEB
\author{S.\,Aoki}\INSTCC
\author{A.\,Ariga}\INSTEE
\author{Y.\,Ashida}\INSTCD
\author{D.\,Autiero}\INSTJ
\author{S.\,Ban}\INSTCD
\author{M.\,Barbi}\INSTE
% ...

\collaboration{The T2K Collaboration}\noaffiliation


\date{\today}

\begin{abstract}
  We report measurements by the T2K experiment of the parameters $\theta_{23}$ and $\Delta m^{2}_{32}$ governing the disappearance of muon neutrinos and antineutrinos in the three flavor neutrino oscillation model. Utilizing the ability of the experiment to run with either a mainly neutrino or a mainly antineutrino beam, the parameters are measured separately for neutrinos and antineutrinos. Using $7.482 \times 10^{20}$ POT in neutrino running mode and $7.471 \times 10^{20}$ POT in antineutrino mode, T2K obtained, $\sin^{2}(\theta_{23})=0.51^{+0.08}_{-0.07}$ and $\Delta m^{2}_{32} = 2.53^{+0.15}_{-0.13} \times 10^{-3}$eV$^{2}$/c$^{4}$ for neutrinos, and $\sin^{2}({\thetabar}_{23})=0.42^{+0.25}_{-0.07}$ and ${\dmsqbar}_{32} = 2.55^{+0.33}_{-0.27} \times 10^{-3}$eV$^{2}$/c$^{4}$ for antineutrinos (assuming normal mass ordering). No significant differences between the values of the parameters describing the disappearance of muon neutrinos and antineutrinos were observed.

\end{abstract}


\maketitle

\section{Introduction}
\label{sec:introduction}
An update to T2K's results on the $\overline{\nu}_{\mu}$ disappearance oscillation analysis\cite{Abe:NumubarPRL} using larger statistics and a substantial improvement to the analysis procedure is presented.
The results presented here include data taken in periods where the beam was operated in neutrino mode, mainly November 2010--May 2013 and in antineutrino
mode, June 2014, November 2014--June 2015, January 2016--May 2016. This corresponds to an exposure of $7.48\times 10^{20}$ and $7.47\times 10^{20}$ protons on target (POT) for neutrinos and antineutrinos respectively, reflecting an increase of $86.3\%$ of the antineutrino mode statistics compared to the result reported in \cite{Abe:NumubarPRL}. Data taken during the same periods were used for the result reported in \cite{Abe:2017uxa}, with the difference that only the muon neutrino and antineutrino candidate events are used for the result presented here. Additional degrees of freedom are also allowed in the present analysis to search for potential differences between the oscillations of neutrinos and antineutrinos.

The standard picture of neutrino oscillations invokes three species of neutrinos and a unitary mixing matrix parameterized by three angles $\theta_{12}$, $\theta_{23}$, $\theta_{13}$ and a CP-violating phase $\delta_{CP}$, plus two mass-squared splittings $\Delta m^{2}_{32}$ and $\Delta m^{2}_{21}$. In this model, the survival probability in vacuum is identical for muon neutrinos and antineutrinos. For the neutrino energies used by T2K, matter effects do not significantly affect this
symmetry. Any difference in the oscillations could be interpreted as possible CPT violation and$/$or evidence of non-standard interactions\cite{Kostelecky:2011gq,Miranda:2015dra}. Non-standard interactions include phenomena not described by the Standard Model (SM).
The analysis presented allows the antineutrino oscillation parameters for $\overline{\nu}_{\mu}$ disappearance to vary independently from those describing neutrino oscillations, i.e., $\theta_{23} \neq \overline{\theta}_{23}$ and $\Delta m^{2}_{32} \neq \Delta \overline{m}^{2}_{32}$, where the barred parameters govern antineutrino oscillations. All other parameters are assumed to be the same for neutrinos and antineutrinos since this data set cannot constrain them. A direct comparison, within the same experiment, of the neutrino and antineutrino oscillation parameters is an important check of this model.

\section{Experimental apparatus}
T2K utilizes the J-PARC facility operating in Tokai, Japan. The neutrino beam illuminates detectors located both off-axis (at an angle of 2.5$^{\circ}$ to the beam axis) and on-axis. The “off-axis” configuration produces a narrow width (in energy) neutrino beam that peaks around 0.6 GeV which reduces backgrounds from higher-energy neutrino interactions.
This is the energy at which the first minimum in the $\nu_\mu$ and $\overline{\nu}_{\mu}$ survival probability is expected to occur at the T2K baseline.
The Super-Kamiokande~(SK) 50-kilotonne water Cherenkov detector\cite{Fukuda2003418,Abe20140211}, situated 295 km away on the off-axis direction, is used to detect the oscillated neutrinos. The detector is divided by a stainless steel structure into an inner detector (ID), which has 11,129 inward-facing
20 inch diameter photomultiplier tubes, and an outer detector (OD),
instrumented with 1,885 outward-facing 8 inch diameter photomultiplier tubes that is
mainly used as a veto. The events at SK are timed using a clock synchronized with the beamline using a GPS system with $<150$ ns timing resolution.

Located 280m from the target are a suite of detectors used to constrain the beam flux and backgrounds. These include the on-axis detector (INGRID\cite{Abe:2011xv}) and a suite of off-axis detectors~(ND280: P$\O$D-$\pi^{0}$ Detector\cite{Assylbekov201248}, FGD-Fine Grained Detector\cite{Amaudruz20121}, TPC\cite{Abgrall201125}, ECAL\cite{ECAL} and SMRD-Side Muon Range Detector\cite{Aoki2013135}). The INGRID is composed of 7 vertical and 7 horizontal modules
arranged in a cross pattern. Its primary purpose is to measure and monitor the beam profile and stability using neutrino interactions. The ND280 off-axis detector is a magnetized composite detector designed to provide information on the $\nu_\mu$ and $\overline{\nu}_{\mu}$ unoscillated spectra directed at SK and constrain the dominant backgrounds. In addition it constrains the combination of flux and interaction cross sections. Details of the experiment can be found in \cite{Abe:2011ks}. 

\section{Analysis description}
The data observed at the far detector are compared to the predictions of the three flavor oscillation model to make statistical inferences. To be able to make those predictions, a model of the experiment is constructed using a simulation of the flux of neutrinos reaching the detectors and a model describing the interactions of neutrinos. The predictions from this model are compared to the data observed in the near detectors to tune the predictions for the far detector by constraining the model parameters. This section describes the different parts of the analysis, focusing on the improvements since the result reported in \cite{Abe:NumubarPRL}.
\subsection{Beam flux prediction}
The fluxes of the different flavors of neutrinos reaching the detectors are predicted by a series of simulations \cite{PhysRevD.87.012001}. The flux and properties of the proton beam reaching the target are measured by the proton beamline monitors, and used as inputs for the simulations. Interactions of the protons in the graphite target and production of secondary hadrons are then simulated using the FLUKA 2011 package~\cite{BOHLEN2014211}. Measurements from hadron production experiments, in particular NA61/SHINE \cite{Abgrall2016}, are used to tune this part of the simulation and the out-of-target interactions. The propagation and decay in flight of the hadrons in the decay tunnel are then simulated using the GEANT3 \cite{GEANT3} and GCALOR \cite{GCALOR} packages. The fluxes are predicted using the same procedure as in \cite{Abe:NumubarPRL}, with updated proton beam parameters (profile of the proton beam on the target) due to the additional data. Several sources of systematic uncertainties (including beamline alignment, hadron production, horn current and proton beam parameters) are considered to produce, for each type of neutrino, an uncertainty on the flux as a function of the neutrino energy. The obtained uncertainties at the peak energy vary between 7\% and 10\% depending on the neutrino flavor, the dominant contribution being the uncertainties on the production of hadrons in the interactions happening in the target. The uncertainties on the hadron interactions occurring outside of the target also have a significant contribution, in particular for the wrong-sign component of the flux (${\nu}_{\mu}$ when running in antineutrino mode, and $\overline{\nu}_{\mu}$ in neutrino mode).

Because of the differences in the production cross-section for positive and negative pions in the proton-carbon interactions in the target, inverting the horn polarities does not simply exchange the neutrino and antineutrino fluxes. The $\overline{\nu}_{\mu}$ flux in antineutrino mode is 20\% smaller than the ${\nu}_{\mu}$ flux in neutrino mode, while the $\nu_{\mu}$ contamination in antineutrino mode is 3.3\% around the peak energy, compared to 2.4\% $\overline{\nu}_{\mu}$ contamination in neutrino mode.

\subsection{Neutrino interaction models}
A significant difference between neutrinos and antineutrinos which needs to be taken into account for a direct comparison of their oscillations is the difference in their interactions with matter. In T2K the signal interaction is the charged current quasi-elastic (CCQE) one, $\nu_\mu +n \rightarrow p +\mu^{-}$ for neutrinos and $\overline{\nu}_\mu +p \rightarrow n + \mu^{+}$ for antineutrinos. For this interaction mode and (anti)neutrinos of 0.6~GeV, the cross section of $\nu_\mu$ on $^{16}$O is larger than that of $\overline{\nu}_\mu$ by approximately a factor of four. The main difference is a result of the difference of the sign of the vector-axial interference term in the cross section~\cite{ref25,ref26}, with additional differences coming from nuclear effects.

Interactions of $\nu$ and $\overline{\nu}$ are
modeled using the NEUT Monte Carlo event generator
\cite{ref20,ref22,ref23}.  CCQE events have been generated according
to the Smith-Moniz Relativistic Fermi Gas (RFG) model~\cite{RFG} with
corrections of long-range nuclear correlations computed in Random
Phase Approximation (RPA)~\cite{ref22}. Multinucleon interaction
(2p-2h) processes have been modeled following~\cite{ref22} and
\cite{ref15}. Single and multi-pion processes are also included with
the same assumptions used in our previous
publications~\cite{Abe:NumubarPRL,Abe:2015awa}.

The initial values and uncertainties of the interaction model
parameters are tuned by a fit of the near-detector data. The fitted values are
used to provide constraints for the fit to extract oscillation parameters of the far detector data. Data from
MiniBooNE~\cite{ref27,ref28} and MINER$\nu$A~\cite{ref29,ref30} on
CCQE-like events are no longer exploited in the near fit for setting
priors for the CCQE axial mass and the normalization of the multi-nucleon
(2p-2h) contribution, but are used in the
choice of the default model; RFG+RPA+2p-2h was chosen because it is
most consistently able to describe current measurements from MiniBooNE
and MINER$\nu$A (see~\cite{ref9} for details).

With respect to our previous disappearance
result~\cite{Abe:NumubarPRL} an additional uncertainty in the
description of the ground state of the nucleus has been
introduced. The difference between the Local Fermi Gas model
implemented in~\cite{ref22} and the Global RFG in NEUT has been
parameterized as a function of lepton momentum and angle and used as an
uncertainty.

The treatment of 2p-2h interactions has also been refined: two
separate, uncorrelated parameters have been introduced for
interactions on C and O in place of an uncertainty on the $A$-scaling
law. This choice is motivated and made possible by the addition of the
water-enriched sample in the near detector fit.  Since part of the
uncertainties on those processes are different for neutrinos and
antineutrinos, an additional 2p-2h normalization factor for
$\overline{\nu}$ was included to supplement these two parameters.

Finally further improvements involve the treatment of coherent $\pi$
production: a reweighting as a function of $E_\pi$ from the Rein-Sehgal model~\cite{Rein:1980wg} to the Berger-Sehgal
one~\cite{Berger:2008xs} was applied to the Monte Carlo. In addition the normalization of this process has been reduced to better match dedicated measurements from
MINER$\nu$A~\cite{Higuera:2014azj} and T2K~\cite{Abe:2016fic}.

\subsection{Near detector analysis}
A binned likelihood fit of the events selected as charged current (CC) interactions in the near detectors is used to constrain the flux and neutrino interaction uncertainties, producing a tuned prediction of the event rates at the far detector. The analysis uses events observed in the tracker (the 2 FGDs and 3 TPCs), with a reconstructed vertex in one of the two FGDs, and identified as a muon neutrino (antineutrino) CC interaction by identifying a $\mu^{-}$ ($\mu^{+}$) using the rate of energy deposition of the particle in the TPCs and the measured momentum in the 0.2 T magnetic field. The events are binned as a function of the momentum and angle of the particle reconstructed as a $\mu^{-}$ or $\mu^{+}$ with respect to the axis of the detector, and arranged in different samples based on the topology of the event observed in the detector. In neutrino beam mode, the samples are made based on the number of pions reconstructed: 0 (enriched in CCQE events), 1 ${\pi^{+}}$ (enriched in CC resonant events) and remaining events (mainly deep inelastic events). In antineutrino beam mode, the samples are based on the number of reconstructed TPC-FGD matched tracks: one (enriched in CCQE events) or more than one (enriched in CC non QE events),  and on whether a $\mu^{+}$ ($\overline{\nu}_{\mu}$ samples) or a $\mu^{-}$ (${\nu}_{\mu}$ samples) was reconstructed. 

Events are further separated according to whether their vertices are reconstructed in FGD1 (CH target) or in FGD2 (42\% water by mass) to give a total of 14 samples \footnote{Target material $\times$ [($\nu$-mode samples )+ ($\overline{\nu}$-mode samples)]
  $\times$ ($\nu$+$\overline{\nu}$)=(water+CH)$\times$[ ( 0$\pi$ + 1$\pi^{+}$ + other)+(1-track+N-track)] $\times$ 2=14  
}. The inclusion of the FGD2 samples reduces the uncertainty on the predictions at the far detector by constraining the parameters specific to oxygen nuclei: nucleon Fermi momentum, nucleon binding energy and the normalization of 2p-2h interactions. The data set for the neutrino beam mode used in the near detector analysis is identical to the previous result ($5.82 \times 10^{20}$ POT), but the statistics in antineutrino beam mode were significantly increased, from $0.43 \times 10^{20}$ to $2.84 \times 10^{20}$ POT, which provides increased ability to constrain the uncertainties in antineutrino running mode, including the ${\nu}_{\mu}$ component of the antineutrino mode beam. Additionally, an improved parameterization of the detector systematic uncertainties was implemented.

There is a total of 651 parameters in the near detector fit, covering flux, interaction and detector uncertainties. The $p$-value, computed by comparing the value of the $\chi^{2}$ obtained when fitting the data to the values obtained for an ensemble of toy experiments, was found to be 8.6\%. The fit also reduces the uncertainties on the expected event rates at the far detector, in particular by introducing anti-correlations between flux and neutrino interactions uncertainties as the near detector measurement is mainly sensitive to the product of the two. The error on the number of expected events in the far detector samples due to these uncertainties is reduced from 10.8\% to 2.8\% for the ${\nu}_{\mu}$ sample, from 11.9\% to 3.3\% for the $\overline{\nu}_{\mu}$ sample, and on the ratio of the expected numbers of $\overline{\nu}_{\mu}$ and ${\nu}_{\mu}$ events from 6.1\% to 1.8\%.

\subsection{Far detector}
The Far Detector employed by T2K is the the Super-Kamiokande (SK) water \v{C}erenkov detector\cite{Fukuda2003418,Abe20140211}.
Events at the far detector (SK) are reconstructed using photomultiplier tube hits chosen based on the arrival time of the hits relative to the leading edge of the neutrino spill.

To construct the analysis samples, events that are fully contained and inside the fiducial volume (FCFV) are selected. Events are defined as fully contained when there is little activity in
the outer detector and as inside the fiducial volume when the distance from the reconstructed interaction vertex to the nearest
inner detector wall is larger than 2 m.
The fiducial mass determined by these criteria is 22.5 kiloton.

In order to enhance the purity of the samples in $\overline{\nu}_\mu$ or $\nu_\mu$ CCQE events,
a single muon-like Cherenkov ring is required, corresponding to a muon momentum
greater than 200 MeV/c, and with no more than one delayed electron.

 The number of data and MC events passing each selection criterion are shown
 in Tables \ref{table:evt_neutrino} and \ref{table:evt_antineutrino}. Expected numbers of events for MC are calculated assuming oscillations in the normal hierarchy scenario with values of the atmospheric parameters corresponding to the result reported in \cite{Abe:2015awa}, $\sin^2(\theta_{23}) = \sin^2(\overline{\theta}_{23}) = 0.528$,
$\Delta m^2_{32}=\Delta \overline{m}^2_{32} = 2.509\times10^{-3}~\mbox{eV}^2/\mbox{c}^4$, and $\sin^2(\theta_{13})= 0.0217$ from \cite{Agashe:2014kda}.
 The fraction of events corresponding to $\overline{\nu}_\mu$ interactions in neutrino beam mode is 6$\%$ while the fraction of $\nu_\mu$ interactions in antineutrino beam mode is 38$\%$.
 The efficiency and purity for $\nu_\mu$ CCQE event selection in the neutrino
mode are estimated to be 71$\%$ and 52$\%$ respectively. For the antineutrino mode
the efficiency and purity are estimated to be 77$\%$ and 35$\%$ for $\overline{\nu}_\mu$ CCQE.
In both modes, the rejection efficiency for NC event is 98$\%$.

\begin{table}
\caption{The number of expected and observed events at SK in neutrino mode
after each selection is applied. Efficiency numbers are calculated
with respect to the number of MC events generated in the fiducial volume (FV interaction).
}
\begin{tabular}{lccccccc}
\hline
\hline
& \multirow{2}{*}{Data\;\;} & Total & \multicolumn{2}{c}{CCQE} & \multicolumn{2}{c}{CCnonQE} & $\overline{\nu}_{e}$+$\nu_e$ \\
&  & MC  &  $\overline{\nu}_{\mu}$ & ${\nu}_{\mu}$ & \;\; $\overline{\nu}_{\mu}$ & $\nu_\mu$ & +NC \\
\hline
FV interaction               &  ---  &  744.9  &  6.4   & 100.2 &  11.6 & 246.1  & 380.6  \\
FCFV                         &  438  &  431.9  &  4.9   & 78.8  &  8.4  & 187.9  & 152.0  \\
Single ring                  &  220  &  223.5  &  4.7   & 73.5  &  4.6  & 70.7   & 70.1   \\
$\mu$-like                   &  150  &  156.6  &  4.7   & 72.2  &  4.4  & 65.6   & 9.6    \\
$P_{\mu}>0.2$~GeV            &  150  &  156.2  &  4.7   & 72.0  &  4.4  & 65.6   & 9.6    \\
$N_{\mbox{decay-e}} <2$      &  135  &  137.8  &  4.6   & 71.3  &  4.1  & 48.5   & 9.2    \\
\hline
   Efficiency ($\%$)         &       &        &  71.9   & 71.2  & 35.3  & 19.7   & 2.4     \\
\hline
\hline

\end{tabular}
\label{table:evt_neutrino}
\end{table}

\begin{table}
\caption{The number of expected and observed events at SK in antineutrino mode
after each selection is applied. Efficiency numbers are calculated
with respect to the number of MC events generated in the fiducial volume (FV interaction).
}
\begin{tabular}{lccccccc}
\hline
\hline
& \multirow{2}{*}{Data\;\;} & Total & \multicolumn{2}{c}{CCQE} & \multicolumn{2}{c}{CCnonQE} & $\overline{\nu}_{e}$+$\nu_e$ \\
&  & MC  &  $\overline{\nu}_{\mu}$ & ${\nu}_{\mu}$ & \;\; $\overline{\nu}_{\mu}$ & $\nu_\mu$ & +NC \\
\hline
FV interaction               &  --- &  312.4 &  30.8  & 20.0  &  38.9  &  74.3  & 148.3 \\
FCFV                         &  170 &  180.5 &  24.9  & 15.0  &  29.1  &  54.1  & 57.2  \\
Single ring                  &  94  &  96.1  &  24.3  & 13.5  &  16.7  &  18.7  & 22.9  \\
$\mu$-like                   &  78  &  74.5  &  24.0  & 13.4  &  16.2  &  17.4  & 3.6   \\
$P_{\mu}>0.2$~GeV            &  78  &  74.4  &  23.9  & 13.4  &  16.2  &  17.4  & 3.6   \\
$N_{\mbox{decay-e}} <2$      &  66  &  68.3  &  23.8  & 13.2  &  15.2  &  12.6  & 3.4   \\
\hline
   Efficiency ($\%$)         &      &        &  77.3  & 66.0  & 39.1  &  17.0  & 2.3   \\
\hline
\hline

\end{tabular}
\label{table:evt_antineutrino}
\end{table}

Table~\ref{tab:oa_syst} summarizes the fractional error on the expected number of SK events using a 1$\sigma$ variation of the flux, cross-section, and far detector uncertainties.

\begin{table*}
\caption{Percentage change in the number of 1-ring neutrino mode and antineutrino mode $\mu$-like events before the oscillation fit from 1$\sigma$ systematic parameter variations, assuming the oscillation parameters $\sin^{2} 2 \theta_{12}=0.846$, $\sin^{2} 2 \theta_{13}=0.085$, $\sin^{2}\theta_{23}=0.528$, $\Delta m^{2}_{32}=2.509\times 10^{-3}~\mbox{eV}^2/\mbox{c}^4$, $\Delta m^{2}_{21}=7.53\times 10^{-5}~\mbox{eV}^2/\mbox{c}^4$, $\delta_{CP} = 0$ and normal hierarchy. The numbers in the parenthesis correspond to the number of parameters responsible for each group of systematic uncertainties. 
}
\begin{tabular}{ l  c  c }
\hline
\hline
\multirow{2}{*}{Source of uncertainty (number of parameters)} & \multicolumn{2}{c}{$\delta n^{\rm{exp}}_{\rm{SK}}$/$n^{\rm{exp}}_{\rm{SK}}$} \\
											      & neutrino mode  & antineutrino mode \\

\hline
Flux+ ND280 constrained cross section (without ND280 fit result)     (61)                & 10.81\%         &           11.92\%	    \\
Flux+ ND280 constrained cross section (using ND280 fit result)    (61)                           & 2.79\%         &           3.26\%	    \\

\hline
Flux+ all cross section                     	         (65)			      & 2.90\%         &         3.35\%   	   \\
Super-Kamiokande detector systematics  (12)                  & 3.86\%     &          3.31\%           \\
Pion FSI and re-interactions 			(12)                 	        & 1.48\%       &	2.06\%			 \\

\hline
  Total (using ND280 fit result)			(77)                           & 5.06\%                  					       &     		5.19\%	 \\
\hline
\hline
\end{tabular}
\label{tab:oa_syst}
\end{table*}

\subsection{Oscillation analysis}
\label{sec:OA}
The analysis method here follows from what was presented in \cite{Abe:NumubarPRL}. As described in Sec.~\ref{sec:introduction} the three flavor neutrino oscillation formalism is extended to include independent parameters $\sin^2({\thetabar}_{23})$ and $\dmsqbar_{32}$ which only affect antineutrino oscillations. Any difference between $\sin^2({\thetabar}_{23})$ and  $\sin^2(\theta_{23})$ or  $\dmsq_{32}$ and $\dmsqbar_{32}$ could be interpreted as new physics. 

With the number of events predicted in the antineutrino sample, the uncertainties on the background models have a non-negligible impact on the measurement of $\sin^2(\thetabar_{23})$ and ${\dmsqbar}_{32}$. The largest is the contribution from the uncertainty on $\sin^2(\theta_{23})$ and $\dmsq_{32}$ due to the significant neutrino background in the antineutrino sample. This provides the motivation for a simultaneous fit of the neutrino and antineutrino data sets.

The oscillation parameters of interest, $\sin^2(\theta_{23})$, $\dmsq_{32}$, $\sin^2({\thetabar}_{23})$ 
and $\dmsqbar_{32}$, are estimated using a maximum likelihood fit to the measured
reconstructed energy spectra in the far detector, for neutrino mode and antineutrino mode $\mu$-like samples. In each case, fits are performed by maximizing the marginal likelihood in the two dimensional parameter space for each pair of parameters.
The marginal likelihood is obtained by integrating over the nuisance parameters $\mathbf{f}$ with prior probability densities $\pi(\mathbf{f})$, giving a likelihood as a function of only the relevant oscillation parameters $\mathbf{o}$:

\begin{equation}
\mathcal{L}(\mathbf{o}) = \int \prod_{i}^{\mathrm{bins}} \mathcal{L}_{i}(\mathbf{o},\mathbf{f}) \times \pi(\mathbf{f}) \,d\mathbf{f},
\end{equation}
where bins denotes the number of analysis bins.
All other oscillation parameters, except $\delta_{CP}$, are treated as nuisance parameters
along with systematic parameters and are marginalized in the construction of the likelihood in accordance with the priors detailed in Table~\ref{tab:osc_params_joint_priors}. $\delta_{CP}$ is fixed to 0 in each fit as it has a negligible impact on the disappearance spectra at T2K. 
Oscillation probabilities are calculated using the full three-flavor oscillation framework~\cite{PhysRevD.22.2718}, with $\sin^2({\thetabar}_{23})$ and $\dmsqbar_{32}$ for $\overline{\nu}$, and $\sin^2(\theta_{23})$ and $\dmsq_{32}$ for $\nu$.
Matter effects, almost negligible in this analysis, are included with a matter density of $\rho$ = 2.6~g/cm$^3$~\cite{Hagiwara:2011}.

\begin{table}
\caption{
Prior constraints of the nuisance oscillation parameters in the fit.
All the Gaussian priors are from \cite{Agashe:2014kda}.
}
 \begin{tabular}{lcc}
 \hline
 Parameter                                               		   		& Prior		& Range    \\
\hline
$\sin^{2}\theta_{23}$                                       				&  Uniform 	& $[0; 1]$      \\
$\sin^{2} 2 \theta_{13}$  								&  Gauss 		& $0.085 \pm 0.005$ 	\\
$\sin^{2} 2 \theta_{12}$  								&  Gauss 		& $0.846 \pm 0.021$ \\
$\Delta m^{2}_{32}$ (NH)              						  &  Uniform		& $[0;+\infty[$ \\
$\Delta m^{2}_{31}$ (IH)  							&  Uniform		& $]-\infty;0]$  \\
$\Delta m^{2}_{21}$      								&  Gauss 		& $(7.53 \pm 0.18) \times 10^{-5}~\mbox{eV}^2/\mbox{c}^4$		\\
$\delta_{CP}$                                               				&  Fixed	 	& 0 \\
\hline
\end{tabular}
\label{tab:osc_params_joint_priors}
\end{table}

Confidence regions are constructed for the oscillation
parameters using the constant $\Delta \chi^2$ method~\cite{Agashe:2014kda}. We define $\Delta \chi^2 = -2 \ln(\mathcal{L}(\mathbf{o})/\mathrm{max}(\mathcal{L}))$ 
as the logarithm of the ratio of the marginal likelihood at a point 
$\mathbf{o}$ in the $\sin^2(\thetapbar_{23})$ -- $\dmsqpbar_{32}$ oscillation
parameter space and the maximum marginal likelihood.
The confidence region is then defined as the area of the oscillation
parameter space for which $\Delta \chi^2$ is less than a standard critical value. This method was used as the difference between the confidence regions produced by it and those obtained using the Feldman-Cousins~\cite{Feldman:1997qc} method was found to be small. For the Feldman-Cousins method, the critical chi-square values were calculated for a coarse set of points in the oscillation parameter space.

\section{Results and discussion}
\begin{figure}
\centering
\includegraphics[width=8cm]{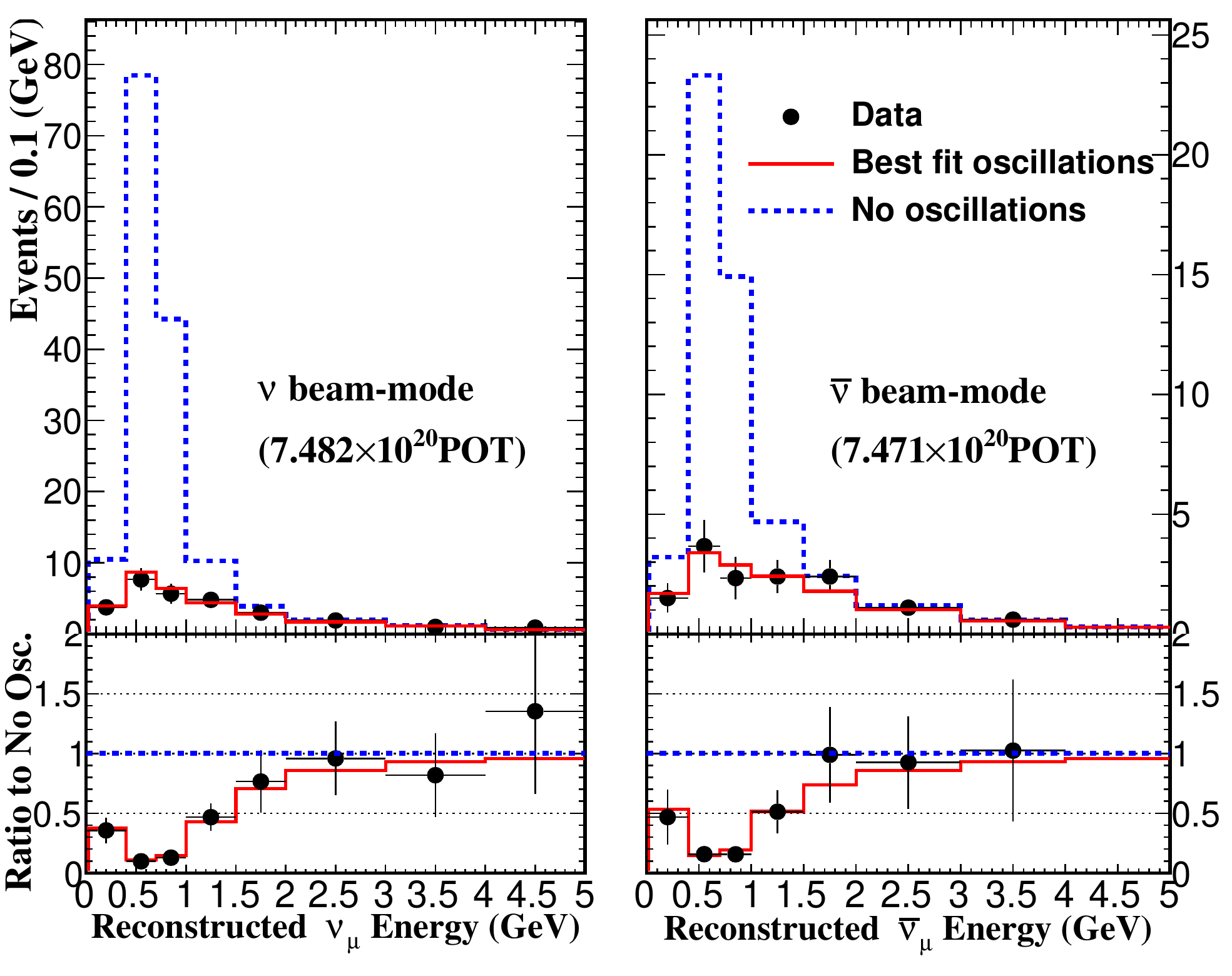}
\caption{Top: Reconstructed energy distribution of the 135 far detector $\nu_\mu$-CCQE candidate events (left) and 66 $\overline\nu_\mu$-CCQE candidate events (right), with predicted spectra for best fit and no oscillation cases.
Bottom:  Ratio to unoscillated predictions.
}
\label{fig:nunb_ratio}
\end{figure}

The reconstructed energy spectra  of
the events observed during neutrino and antineutrino running modes are shown in
Fig. \ref{fig:nunb_ratio}. These are overlaid with the predictions for the best fit values of the oscillation parameters assuming normal hierarchy, and in the case of no oscillations.
The lower plots in Fig. \ref{fig:nunb_ratio} show the ratio of data to the unoscillated spectrum.

\begin{figure}
\centering
\includegraphics[width=8cm]{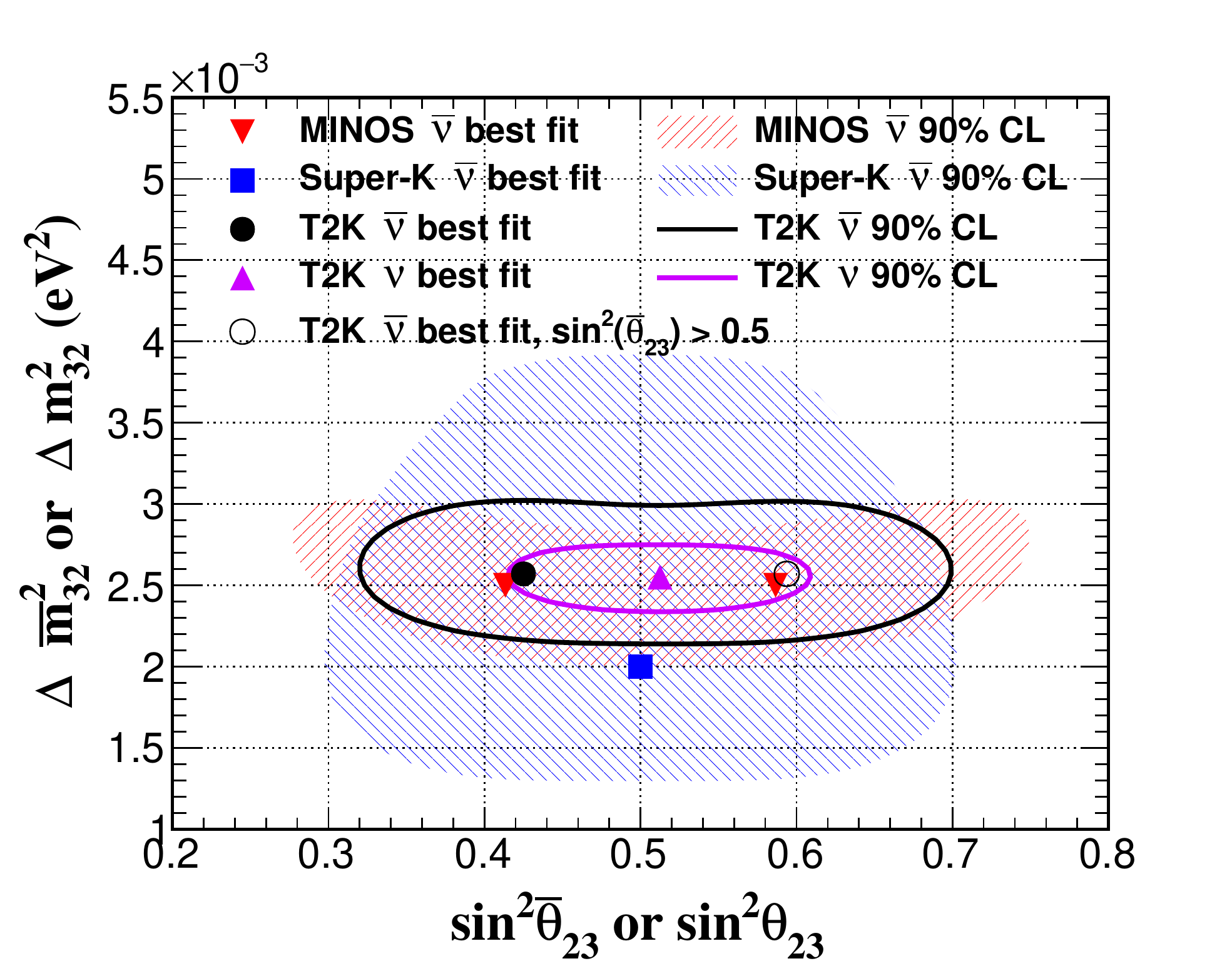}
\caption{90\% confidence regions for $\sin^2(\theta_{23})$ and $\dmsq_{32}$ in $\nu$ mode (corresponding to 7.482$\times10^{20}$POT) and $\overline\nu$-mode (corresponding to 7.471$\times10^{20}$POT). Normal hierarchy is assumed. 90\% confidence regions obtained by SK ~\cite{Abe:2011ph} and MINOS~\cite{Adamson:2012rm} for $\overline\nu$  are also shown. The best fit in the case $\sin^2(\theta_{23})>0.5$ is also displayed for comparison with the MINOS result.
}
\label{fig:final_contour}
\end{figure}

Assuming normal hierarchy, the best fit values obtained for the parameters describing neutrino oscillations
are $\sin^2(\theta_{23}) = 0.51$ and $\dmsq_{32} = 2.53\times10^{-3}$eV$^{2}$/c$^{4}$
   with 68\% confidence intervals of 0.44 -- 0.59
and 2.40 -- 2.68 ($\times10^{-3}$eV$^{2}$/c$^{4}$) respectively.
For the antineutrino parameters, the best fit values are
$\sin^2(\thetabar_{23}) = 0.42$ and
$\dmsqbar_{32} = 2.55\times10^{-3}$eV$^{2}$/c$^{4}$ with 68\%
confidence intervals of 0.35 -- 0.67
and 2.28 -- 2.88 ($\times10^{-3}$eV$^{2}$/c$^{4}$) respectively. For comparison, the best fit values (68\% confidence intervals) obtained when using the same oscillation parameters for neutrinos and antineutrinos are 0.52 (0.43 -- 0.595) for $\sin^2(\theta_{23})$ and 2.55 (2.47 -- 2.63) $\times10^{-3}$eV$^{2}$/c$^{4}$ for $\dmsq_{32}$.
The values for the inverted hierarchy can be obtained by replacing $\dmsqpbar_{32}$ by $-\dmsqpbar_{31}$, effectively changing the sign of $\dmsqpbar_{32}$ and shifting its absolute value by $-\dmsq_{21}=-7.53\times 10^{-5}~\mbox{eV}^2/\mbox{c}^4$.\\
A goodness-of-fit test was performed by comparing the best fit value of the $\chi^{2}$ to the values obtained for an ensemble
of toy experiments generated with systematic variations and statistical fluctuations,
giving a $p$-value of 96\%. In Fig.~\ref{fig:final_contour}, the 90\% confidence regions obtained for the parameters describing the disappearance of muon antineutrinos are compared to the
corresponding measurements by the Super-Kamiokande collaboration using atmospheric
antineutrino data~\cite{Abe:2011ph} and the MINOS collaboration using beam
antineutrino data~\cite{Adamson:2012rm}. This new measurement is consistent with the results obtained by the SK and MINOS collaborations.

Our new measurements of [$\sin^2(\theta_{23})$, $\Delta m_{32}^2$] and
[$\sin^2(\thetabar_{23})$, $\dmsqbar_{32}$], using neutrino mode data corresponding to 7.482$\times10^{20}$ POT
and antineutrino mode data corresponding to 7.471$\times10^{20}$ POT, provide
no indication of new physics. When analyzed both in the normal and inverted hierarchy hypotheses the results are consistent with the expectation that the parameters describing the disappearance of muon neutrinos and antineutrinos are equivalent. The data related to this measurement can be found in \cite{DataRelease}.

\begin{acknowledgments}
We thank the J-PARC staff for superb accelerator performance. We thank the
CERN NA61/SHINE Collaboration for providing valuable particle production data.
We acknowledge the support of MEXT, Japan;
NSERC (Grant No. SAPPJ-2014-00031), NRC and CFI, Canada;
CEA and CNRS/IN2P3, France;
DFG, Germany;
INFN, Italy;
National Science Centre (NCN) and Ministry of Science and Higher Education, Poland;
RSF, RFBR, and MES, Russia;
MINECO and ERDF funds, Spain;
SNSF and SERI, Switzerland;
STFC, UK; and
DOE, USA.
We also thank CERN for the UA1/NOMAD magnet,
DESY for the HERA-B magnet mover system,
NII for SINET4,
the WestGrid and SciNet consortia in Compute Canada,
and GridPP in the United Kingdom.
In addition, participation of individual researchers and institutions has been further
supported by funds from ERC (FP7), H2020 Grant No. RISE-GA644294-JENNIFER, EU;
JSPS, Japan;
Royal Society, UK;
the Alfred P. Sloan Foundation and the DOE Early Career program, USA.
\end{acknowledgments}

\vfill

\bibliography{Numubar}

\end{document}